# Measurements of energy spectra of relativistic electrons and gamma-rays avalanches developed in the thunderous atmosphere with Aragats Solar Neutron Telescope

A. Chilingarian[1,2], G. Hovsepyan[1,2], T. Karapetyan[1,2], B. Sargsyan[1,2], S. Chilingaryan[1]

[1]*A. Alikhanyan National Lab (Yerevan Physics Institute), Yerevan 0036, Armenia*

[2] *HSE University, Moscow 109028, RF*

[3]*National Research Nuclear University MEPhI, Moscow 115409, RF*

**Abstract**

Aragats solar neutron telescope (ASNT) is a unique instrument allowing to measure the energy spectra of electrons accelerated and multiplied in the strong electric fields of the atmosphere.
We describe the instrument setup, its operation condition, software, and hardware triggers. We present energy spectra of a very large thunderstorm ground enhancement (TGE) event observed on 6 October 2021. The detector response function, algorithm to recover energy spectra from the energy release histograms also are presented. The spectra recovery procedure is verified by simulation of the response function of the SEVAN detector, operating nearby ASNT. SEVAN is a stacked 3 layered detector, interlayered by lead filters registering both charged and neutral species of cosmic rays. The simulated and measured count rates of all 3 layers of the SEVAN detector show good agreement within 20%.

1. Introduction. Particle detectors in High-energy atmospheric physics

Around 100 years ago B.F.J.Shonland advised and encouraged by C.T.R. Willson, one of the first particle physicists and leading researchers of atmospheric electricity, had designed and commissioned particle detectors for registering enigmatic "runaway" electrons nearby the violent South African thunderstorms [1,2]. However, due to an erroneous model they were looking for GeV runaway electrons, which, as we know nowadays never can reach such energies due to catastrophic energy losses by radiation processes, which are predominant at energies above a few tens of MeV. Thus, used ionization chambers, Geiger-Muller counters, and electrometers fail to measure the electron flux during thunderstorms [3,4].
After a half-of century, enigmatic runaway particles (thunderstorm ground enhancements, TGEs [5]) were detected in the troposphere by balloon flights, by scintillation and NaI detectors on aircraft, in space, and finally on mountain altitudes by NaI detectors located on roofs of nuclear power stations in Japan, by facilities of cosmic ray station on Aragats and by East-European network of SEVAN detectors (see reviews of measurements and instruments used in [6-7]). Thus, finally was established a multi-sensory observatory using various particle detectors, systematically measuring fluxes of gamma rays, electrons, muons, and neutrons of atmospheric origin, their energy spectra, and correlations with the near-surface electric field, and lightning occurrences [8]. These measurements are supplemented by meteorological and optical observations. In addition, the measurements performed by several research groups at various observation sites with similar detectors confirmed the main parameters of the thunderstorm ground enhancements, surplus fluxes of electrons, gamma rays, and neutrons during thunderstorms [9,10].

The most difficult and most important experimental challenge was the measurements of the electron content of TGEs. Electrons are attenuated very fast due to ionization losses in the air and reach the ground in amounts that made impossible their reliable identification against the huge background of gamma rays, which exceeds the electron fluxes by 1-2 orders of magnitude. Certainly, small NaI spectrometers cannot recover weak electron fluxes, we need much larger spectrometers with the possibility to distinguish between charged and neutral particles. The only spectrometer operated for the High-energy atmospheric physics (HEPA), which routinely measures electron and gamma ray spectra is the Aragats Solar Neutron Telescope (ASNT, [11]). ASNT was a part of the worldwide network aimed to detect neutrons born in photosphere and reach Earth bringing direct information from the sun. The network is coordinated by the Solar-Terrestrial laboratory of the Nagoya University [12, 26] and consists of seven detectors of the same type located in different countries around the globe to observe the sun 24 hours a day. The ASNT was the most advanced among these detectors, allows measurements of energy spectra and solving many other physical problems. After finishing the solar physics program, ASNT was used for key measurements providing deeper understanding of the origin of TGE [13] and the structure of accelerating electric field in the lower atmosphere. One of the most important tasks in the high-energy physics in the atmosphere (HEPA) is to gain insight into the origin of the enhanced particle fluxes, the so-called, thunderstorm ground enhancements (TGEs) registered by surface detectors during thunderstorms. The key point in the TGE study is the measurement of the energy spectra of avalanche electrons that reach the ground. Electrons are attenuated very fast after exiting from the strong accelerating atmospheric electric field due to ionization losses, and only a weak intensity flux can reach the ground. Usually, the height of the strong atmospheric field is above 150 m and it makes separation of electron flux from the 10-100 times more abundant gamma ray flux practically impossible. However, in very rare cases, a few times per year, the electric field on Aragats can go down below 100 m for a few minutes, and for these rare TGEs, the ASNT spectrometer allows us to detect and recover electron and gamma ray fluxes separately.

2. **Aragats Solar Neutron Telescope (ASNT)**

The Aragats Solar Neutron Telescope (ASNT) was deployed in 2003. In addition to detecting neutrons born in violent flares on the Sun, the ASNT also has the possibility to detect charged fluxes and measure differential energy spectra of electrons and gamma rays originated in avalanches developed in the thunderous atmosphere. ASNT, as well, monitor high-energy muon flux traversing detector horizontally within azimuthal angles of 0-0.5, and 0-3 degrees.

ASNT is formed by 4 separate identical modules, as shown in Fig. 1. Each module consists of forty 50 x 50 x 5 $cm^3$ scintillator slabs stacked vertically on a 100 x 100 x 10 $cm^3$ plastic scintillator slab. Scintillators are finely polished to provide good optical contact of the assembly. The slab assembly is covered by white paper from the sides and bottom and firmly kept together with special belts. The total thickness of the assembly is 60 cm. Four scintillators of 100 x 100 x 5 cm³ each are located above the thick scintillator assembly to indicate charged particle traversal and separate the neutral particles by "vetoing" charged particles (the efficiency to detect neutral particle by 5 cm thick scintillator is ≈5%). A scintillator light capture cone and Photo Multiplier Tube (PMT) are located on the top of the scintillator housings.

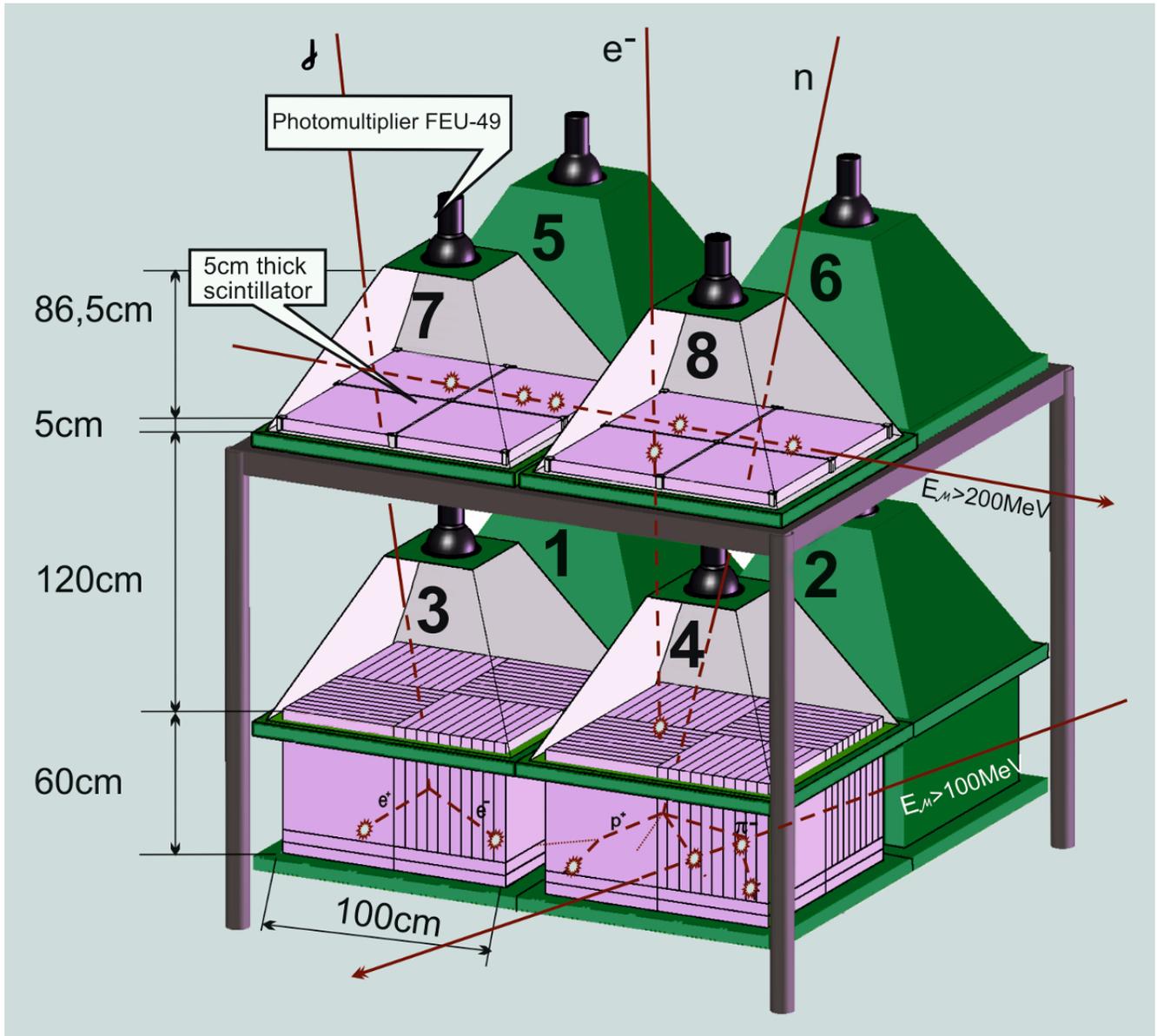

**Figure 1. Assembly of ASNT with a schematic view of the different particles traversing the detector. TGE electrons are registered by both upper and lower layers; gamma rays and neutrons – by invoking the veto option (no signals from the upper scintillators), horizontal muons – by the condition of operation of 2 upper scintillators from 4 and no signal in the lower scintillators (to prevent registration of EAS events) and by very large energy release.**

The main ASNT trigger stores the analog signals (PMT outputs) from all 8 channels if at least one channel reports a signal above the preset threshold (we use several thresholds, equivalent to the traversal of 4 MeV (after May 2018), or 7- MeV particles to avoid the registration of the Radon progeny gamma radiation). Additional, so-called, software triggers, utilize the whole bulk of the information from amplitude-digital-converters (ADC) on the energy releases in 8 scintillators. The software triggers are not fixed in electronics and it is possible to change them remotely. The list of the most important software triggers is as follows:

1. 2-s time series of count rates of all 8 channels of ASNT (previously was used 1-minute and 10-s time series);

2. Count rates from different incidence directions, 16 possible coincidences of 4 upper and 4 bottom scintillators correspond to 9 solid angles for different azimuthal and zenith directions. This mode is important to show that during thunderstorm, particle fluxes come only from near-vertical direction (coincidences 1-5, 2-6, 3-7, 4-8), other directions never show any enhancements;

3. Count rates of the special coincidences. Most important from all possible coincidences are ones, which separate neutral and charge particles ("01" and "11", where 0 – denotes no signal in the ASNT layer, 1 is only one signal in the 4 scintillators of each layer), also very important are "MANY-MANY", and "0-MANY" coincidences, where "MANY" denotes more than 1 signal in the layer. "MANU-MANY" denotes a very rare TGE, when electrons are traversing both upper and lower layers, "0-MANY" – only avalanche of gamma rays reach the detector (the efficiency to register gamma ray in the thin upper scintillators is ≈5%, and in thick scintillators below reaches ≈80% at high energies);

4. 10-s histograms of energy releases in every of 8 channels of ASNT;

5. The same as in the previous point with invoking veto option (no signal in the upper layer). Subtracting histograms, we obtain separate histograms related to neutral and charge particle energy releases.

6. Inter-channel correlation matrices: For each minute, 5-s count rates of 8 ASNT channels are calculated and passed to the analysis software. The sequence of 5-s data is gathered for 1 minute for a total of 12 number strings. Then 8 x 8 correlation matrix is calculated and stored in the database for each minute. The correlation matrix provides a test to spurious signals in one or more detector channels, and the channel crosstalk. The enhancement of the count rate in the detector due to TGE should be accompanied by the coherent enhancements of correlations between vertically arranged scintillators.

All data is in XML format, permitting addition of metafiles with detailed information about detector operation conditions and other necessary information. In this way, we secure the multiyear operation during which several detector parameters are changed. Each minute the data is transferred by wireless connections to CRD headquarters in Yerevan and stored in the MySQL database. Online data is available from HTTP servers at Yerevan and Nor Amberd, as well as at the mirror server in Germany. Analysis facilities are provided for users via the ADEI multivariate analysis platform with reach possibilities of multivariate visualization, correlation analysis, and many others [14].

The schemes and operation characteristics of data acquisition (DAQ) electronics are described in [15]. The Data control system (DCS) of ASNT consists of programmable high voltage power supply for PMT, RS-485 local network, microcontroller, and mini-PC. The PMT, preamplifier, and HV supply are placed in the metal case on the top of the PMT housing. The rest of the electronics is placed inside the ASNT electronics board. The output pulses from the preamplifier are fed through coaxial cables to the Logarithmic ADCs inputs. The principle of logarithmic ADC operation is based on the measurement of the decay time of oscillation in the parallel RLC circuit with a well-known Q-factor. ADC output signals are sequences of NIM (Nuclear Instrumentation Module) standard pulses with ~1 MHz frequency and the number of pulses is proportional to the logarithm of the area (charge) of the measured current pulse.

The operational frequency of electronics is set to ≈10 kHz. In Fig.2 we show time series of count rates of 8 ASNT channels. We use now 2-s time series for obtaining the fine structure of TGE including its abrupt termination by lightning flash. As we can see in the inset to Fig. 2 the thick and thin scintillators operate stable and relative errors of the 2-s time series are rather small. As we can see from 1.5 hours operation on 6 October 2021, the background frequency was 11,929 for 2 seconds, i.e., the frequency of triggers was ≈ 6 kHz. Thus, the instrument is capable to measure additional neutron flux from the sun, which is expected to never exceed 10 Hz. However, TGEs at maximum can exceed background more than twice, thus the maximal operation frequency can be reached and in expectational cases, some seconds can be lost.

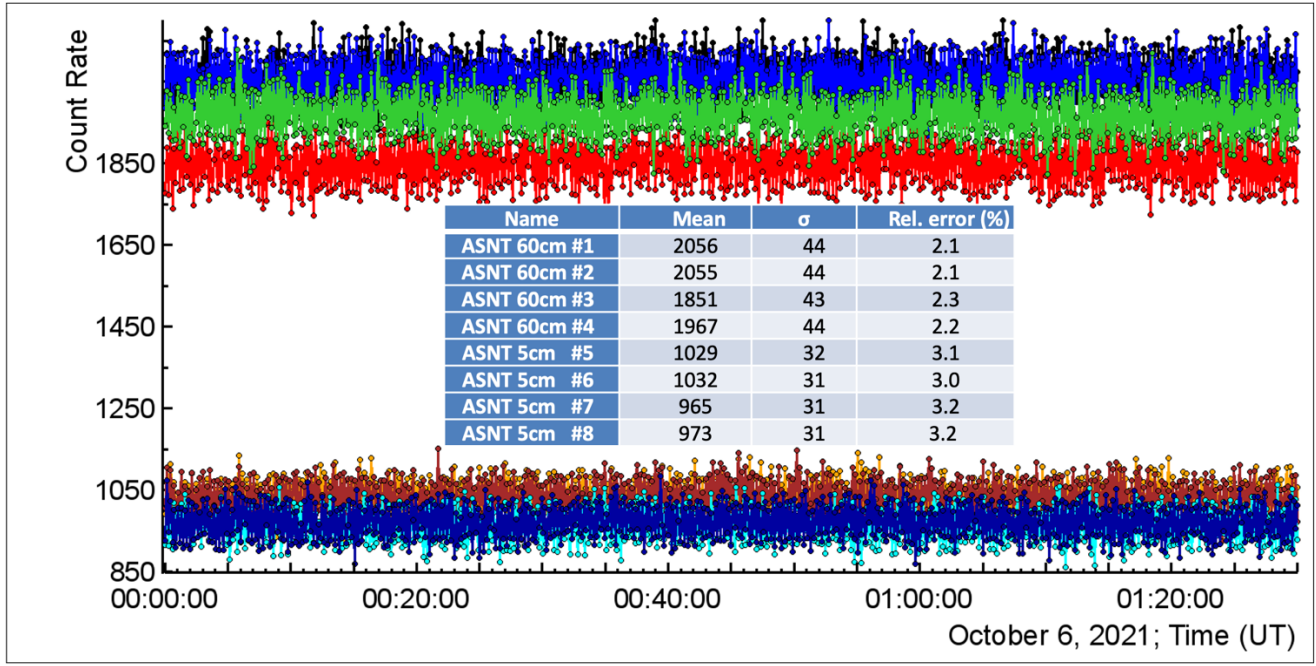

**Figure 2. Two-second time series of 4 5-cm thick plastic scintillators (5-8, see Fig.1, upper "veto" layer of ASNT detector) measured just before large TGE event. In the inset mean values, mean square deviations, and relative errors of each channel.**

In Fig. 3 we show a large TGE with the duration of 3 minutes (2:01 – 2:04) occurred just after the fair-weather period shown in Fig.2. TGE was abruptly terminated by a lightning flash in the end of its development. We use the mean values and mean square deviations computed before TGE for the calculation of significance of the TGE peaks measured in gamma ray and electron fluxes. In Fig. 3a we show the enhancement (in percent to fair-weather value) of count rate of ASNT detector measured by the upper and lower scintillators of ASNT.

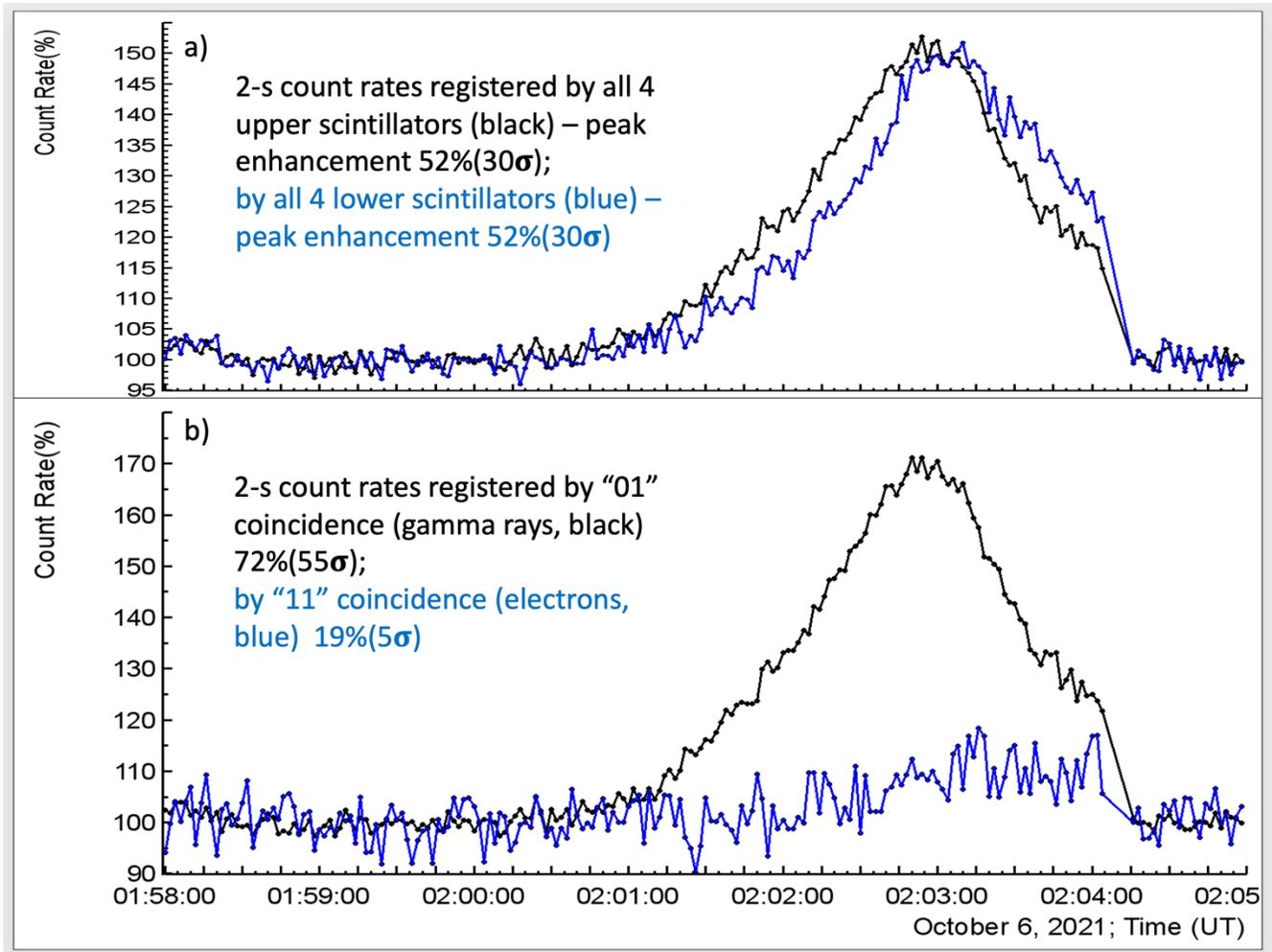

**Figure 3. Parameters of the large TGE abruptly terminated by a lightning flash: a) Count rate enhancements registered by the upper and lower layers of the detector; b) Count rate enhancement of electron and gamma rate fluxes. The percent of the enhancement of 2-s time series are given relative to the fair-weather background (Fig.2).**

As we can see from Fig.3b the significance of the gamma ray count rate peak is much larger compared with the electron peak due to the very fast attenuation of electrons in the air after exiting from the region of strong accelerating atmospheric electric field. However, the electron content is rather large reaching enhancement of ≈ 20% at the end of TGE (2:03 – 2:04 UT), thus making it possible to recover separately energy spectra of electrons and gamma rays, the most important parameters of the TGE.

In is interesting to note, that TGE was accompanied by bright lights in the skies above the station seen in the shots of panoramic cameras, see Fig. 4a and 4b.

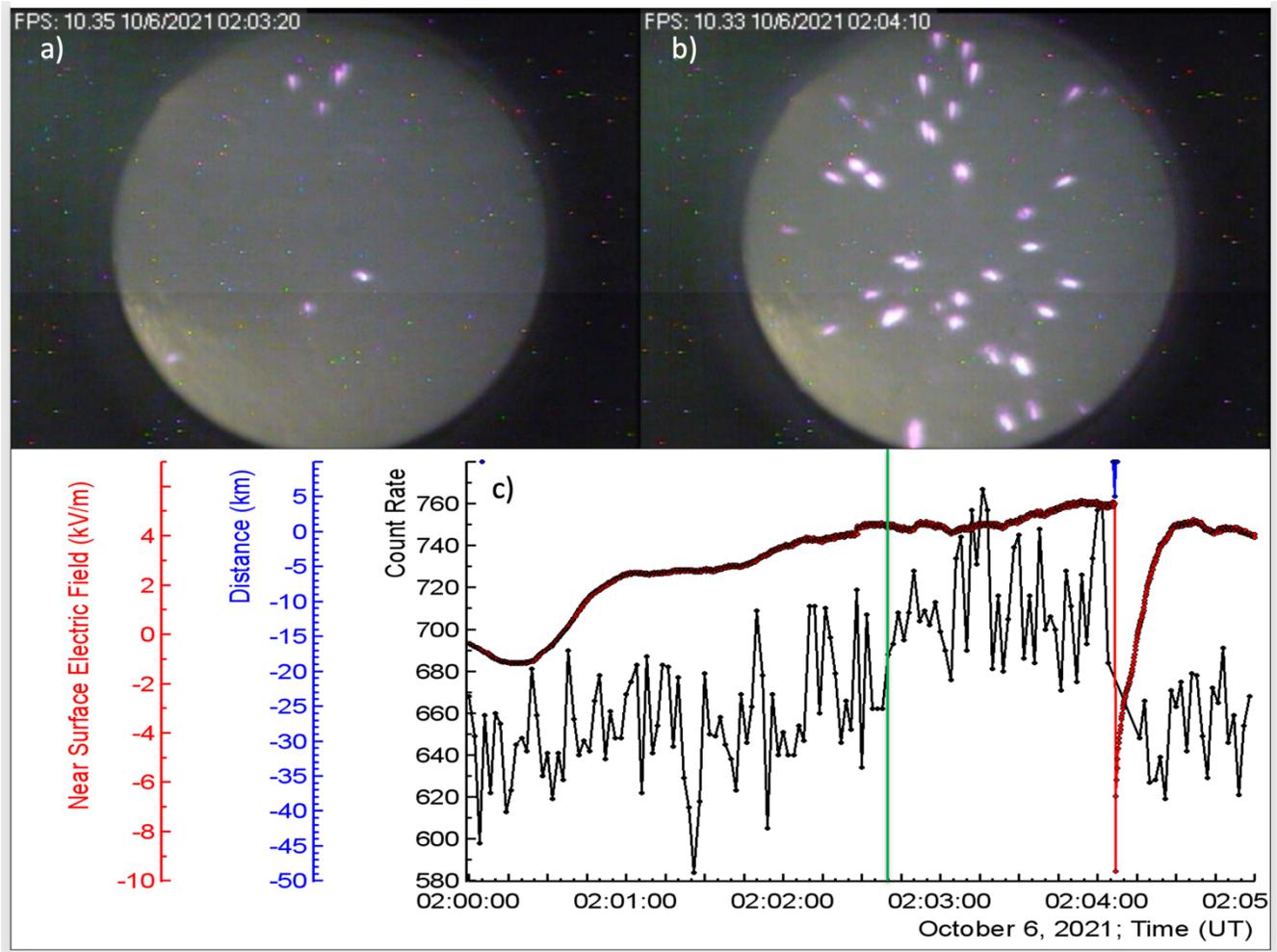

**Figure 4.** Lights were detected above research station a) and b) at maximum electron flux denoted by black curve in c) between green and red lines (2:02:50-2:04:10). By red curve is denoted the near surface electric field measured with frequency 20 Hz by EFM-100 electric field sensor produced by the BOLTEK company. The abrupt decrease of the electric field is due to the nearby lightning flash (1.8 km, denoted by blue line).

In the next sections we will describe the detector response calculation, the procedure of energy spectra evaluation, and performed tests to ensure the correct recovery of energy spectra.

## 3. Recovery of the energy spectra of TGE electrons and gamma rays

Modeling of the ASNT response was performed with the GEANT4.10 package [16]. The structure of detector material and the building construction were described in all details. In the simulation, the building slab and detector tube are described as solid absorbers with averaged thickness: iron structures are represented as a solid iron sheet with 3.5 mm thickness, wooden slabs as a whole absorber with 10mm thickness, slate slab with 5 mm thickness, and detector tube as an iron sheet with 1.5 mm thickness. In the simulation were included fluctuations of light collection in detectors with pyramidal construction ~25% [17] and the absorption in 60-cm scintillators of ASNT, which depends on the depth of particle trajectory in scintillator [18].

Secondary particle fluxes used for the background simulation were calculated with the EXPACS WEB calculator [19,20]. The background was determined by the overall number of secondary cosmic rays at the observation level, including neutron, proton, alpha particle, muon, electron, positron, and gamma ray fluxes. The zenith angle dependence of the incident particles intensity was $Cos^{2.5}(\theta)$ and azimuthal dependence was assumed to be uniform from 0 to $2\pi$. For each detector, the 1-minute histograms of energy release amplitudes were obtained. The agreement of the simulation with the experiment was checked by comparing the position of the maximum of ionization losses. For this purpose, the simulated energy release in the detector was converted to LADC codes [21] by the expression:

$$k = d * \ln(E_{dep}/E_0) + k_o \qquad (1),$$

where $k \geq 1$ is an integer part of the expression on the right, $d = 10.5$ is the LADC scale factor, Edep is the energy release in the plastic scintillator, $E_0 = 10.75$ MeV is the average muon energy release in a 5 cm thick scintillator and ko is the LADC code corresponding to the maximum ionization loses in the detector. As can be seen from (1) the value of ko determines the threshold value of $E_{dep}$ at which $k = 1$.

In Fig. 5 we see a good agreement of the simulated distribution of ADC codes for the 60 cm detector in comparison with measured. The large enhancement of the experimental distribution in the low energy region, from 1 to10th LADC codes (corresponding to energy releases below 2 MeV) is due to the contribution of gamma radiation of Radon progeny abundant on Aragats, which was not included in the simulation. Thus, we keep the energy threshold of recovered energy spectra well above 2 MeV to avoid distortion of spectra.

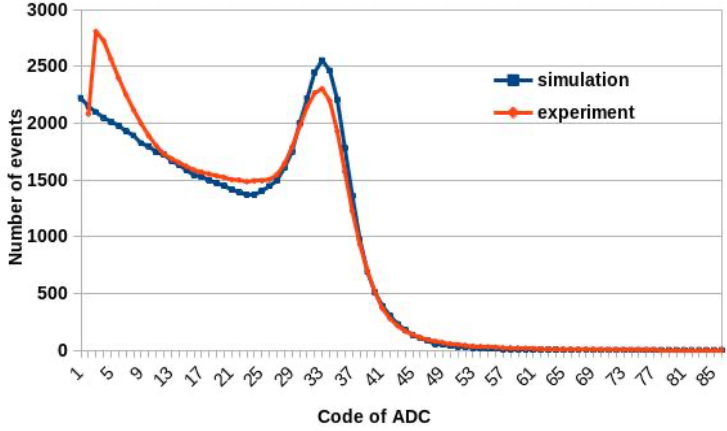

**Figure 5. Simulated and measured histograms of LADC codes for the 60 cm scintillator.**

As cosmic ray spectra are very steep, thus, the correction procedure for the bin-to-bin migration is vital for genuine spectra. The response function that allows the correction of spectra is mathematically expressed as a first-class Fredholm equation:

$$f(x) = \int W(x,y)\varphi(y)dy \qquad (2),$$

where $f(x)$ is the measured distribution of a physical quantity, $\varphi(y)$ is true distribution and $W(x,y)$ is the transformation matrix. It is possible to recover the "true distribution" by solving the inverse problem, of cosmic rays by using the measured distribution $f(x)$ and discreet form of the transformation matrix $A_{i,k}$:

$$x_k = A_{i,k}\, y_i, \quad i,k = 1, N. \qquad (3),$$

where $x_k$ are measured intensities, $y_i$ are its true values, and N number of bins. The $A_{ik}^{-1}$ coefficients had been determined by solving the direct problem of cosmic rays i.e., by simulation of the particle traversal through the detector with GEANT4.10 package [16]. The energy deposit $E_{dep}$ (MeV) in the spectrometer was simulated for a given energy of particle $E_i$ (i=1,50) covering energy range from 1 do 100 MeV. Conversion of energy to LADC codes was done according to Eq. (1), where $E_0$=10.75 MeV and $K_0$ is the code corresponding to the maximum muon energy deposit in the 5cm scintillator; $10^5$ events had been simulated for each $E_i$. The obtained in this way matrix $B_{i,k}$ contains the number of particles with energy $E_i$ detected with a code k (k=1,..,50). The matrix represents the probabilities for the "true" particle energies $E_i$ to be measured with LADC code k. The larger or smaller the code relative to the expected for the fixed energy value, the smaller number of particles fall in these bins. However, due to large fluctuations, the width of the code distribution is rather wide. This procedure uses particle registration efficiency as a function of energy. The transition from $B_{i,k}$ to $A_{ik}^{-1}$ matrix is done by normalizing it in accordance with a total number of simulation trials. When normalizing, the *a priori* information on the recovered spectrum had been taken into account [22]. Finally, from the obtained discreet intensities recovered in 130 bins, the differential energy spectra of gamma rays and electrons were approximated by the power function

$$dJ/dE = AE^{-\gamma} \qquad (4).$$

The parameters of the spectrum were determined by minimizing $\chi^2$ function with CERN MINUIT code [23] for each minute of TGE, see Fig. 6 and Table 1.

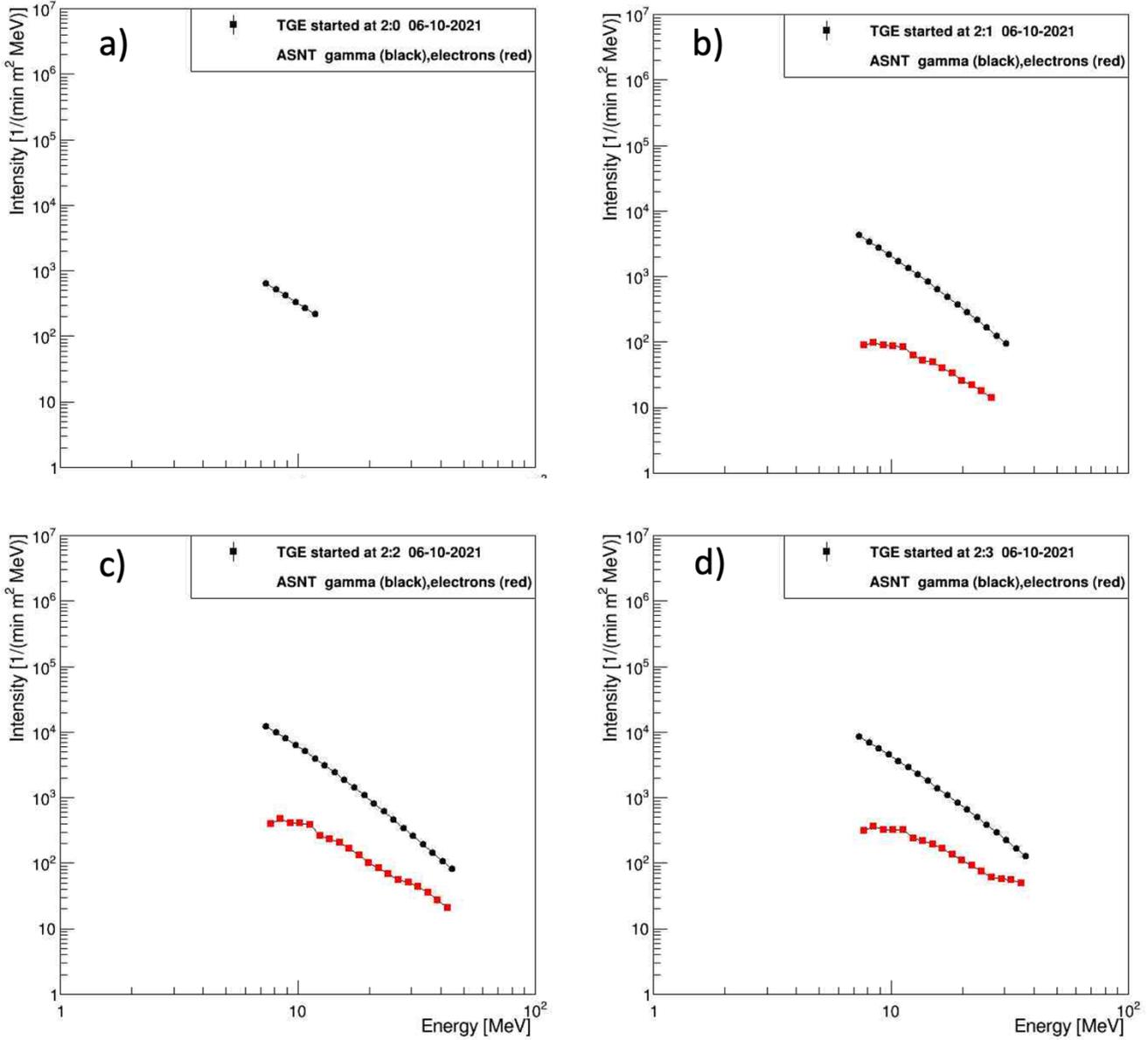

**Figure 6. Recovered differential energy spectra of 4 minutes of TGE**

In Table 1, we can see that electron share in TGE is enlarging (in agreement with Fig. 3) till lightning terminates the TGE. Electron flux produces enough ionization to make a path for the lightning leader moving in direction to the earth's surface.

**Table 1. Recovered integral intensities and electron-to-gamma ray ratios for the TGE particles with energies larger than 7 MeV**

|  | Integral intensities (count rates) ≥7MeV | | Ne/Nγ (%) |
| --- | --- | --- | --- |
| Date and Time | Electrons | Gamma rays | |
| **10.06.2021, 2:00-2:01** | - | 3.739E+03 | |
| **10.06.2021, 2:01-2:02** | 1.494E+03 | 2.406E+04 | 6.3 |
| **10.06.2021, 2:02:2:03** | 5.994E+03 | 7.320E+04 | 8.2 |
| **10.06.2021, 2:03:2:04** | 6.579E+03 | 5.117E+04 | 12.9 |

To check the recovered spectra, we use them for simulation of the count rates of another independent detector, located nearby ASNT, namely SEVAN 3-layered detector, a unit of East-European network [24]. SEVAN is a stacked detector of 5 cm thick scintillators overviewed by photomultipliers similar to ASNT single unit. However, scintillators are interlayered by lead filters, and electronics counts all possible coincidences of layer "firing". Table 2 shows the observed and simulated averaged count rates for the three SEVAN scintillators as well as for the two coincidences, which are selected charged particles ("100") and neutral particles ("010") for the minute of the maximum flux and for background for 10 minutes before TGE start.

**Table 2. Measured and calculated count rates of SEVAN layers and special coincidences at fair-weather (background) and during TGE**

| | Upper scintillator. Measured (left) and simulated | | Middle scintillator. Measured (left) and simulated | | Lower scintillator. Measured (left) and simulated | | "100" coincidence Measured(left) and simulated | | "010" coincidence Measured (left) and simulated | |
| --- | --- | --- | --- | --- | --- | --- | --- | --- | --- | --- |
| **Background** | 30296 | 30479 | 7983 | 8399 | 17805 | 15549 | 21146 | 19835 | 2387 | 2360 |
| **TGE** | 10450 | 8810 | 691 | 798 | 383 | 284 | 10619 | 8860 | 697 | 820 |

As seen from Table 2, there is a rather good agreement of measured and simulated count rates, obtained by applying the recovered with ASNT energy spectra to SEVAN detector; measured and simulated count rates coincide within ±20%, which is quite satisfactory for the cosmic ray experiments with inherent uncertainties and large fluctuations of almost all parameters of the incident particle fluxes.

## Conclusions

Electrons from relativistic runaway avalanches (RREA) very rarely reach the ground and be registered. Due to its unique design ASNT is capable to separate gamma ray and electron fluxes and measure energy spectra of both species of the RREA. Energy spectra measured by the ASNT detector provide the key evidence for proving the model of TGE as a ground "mapping" of the electron-gamma ray avalanche developed in the thunderous atmosphere.

The detector response calculation and algorithm for inverse problem solving was applied to the thunderstorm ground enhancement that occurred on 6 October 2021.

The cosmic ray background, electron, and gamma ray energy spectra were recovered and tested by calculation of count rates of the SEVAN detector. The experimentally observed and simulated particle fluxes at the maximum of TGE development agree with each other with 20% accuracy.

A strong electric field is extending very low above the ground, which is demonstrated by the enhancement of the "11" coincidence at the minute 2:03-2:04 UT (Fig. 3b), by proximity of the gamma ray and electron energy spectra at the same minute (Fig. 6d) and by the lights in the skies appeared (Fig. 4a and 4b).

## Acknowledgments

We thank the staff of the Aragats Space Environmental Center for the uninterruptable operation of all particle detectors and field meters, authors thank Danielyan Varugan for designing of the ASNT DAQ electronics. The authors acknowledge the support of the Science Committee of the Republic of Armenia (research project № 21AG-1C012), in the modernization of the technical infrastructure of high-altitude stations. We also acknowledge the support of the Basic Research Program at HSE University, RF for the assistance in running facilities registering atmospheric discharges.